\documentclass[10pt,a4paper]{article}
\usepackage[utf8]{inputenc}
\usepackage[english]{babel}
\usepackage{url}
\usepackage{hyperref}
\usepackage{amsmath}
\usepackage{amsfonts}
\usepackage{amssymb}
\usepackage{graphicx}
\usepackage{float}
\usepackage{lipsum}
\usepackage{multicol}
\usepackage{epstopdf}
\usepackage{xcolor}
\usepackage{enumitem}
\usepackage{tabularx,booktabs,makecell}
\usepackage{algorithm}
\usepackage{cite}
\usepackage{amsmath,amssymb,amsfonts}
\usepackage{algpseudocode}
\usepackage{textcomp}
\usepackage{xcolor}
\usepackage{blindtext}
\usepackage{subfig}
\usepackage{cuted}
\usepackage{physics}
\usepackage{xcolor}
\usepackage[toc,page]{appendix}
\usepackage[long]{optidef}
\usepackage[detect-none]{siunitx}
\usepackage{multirow}
\usepackage{array}

\definecolor{azul}{rgb}{0.0, 0.53, 0.74}
\usepackage[left=2.00cm, right=2.00cm, top=2.00cm, bottom=2.00cm]{geometry}
\author{Aprendiendo \LaTeX\,}
\title{Mi primer paper}

\begin{document}
	\vspace{7.5mm}
	
	\begin{center}
		{\Large \textbf{Edge-to-Cloud Computations-as-a-Service in Software-Defined Energy Networks for Smart Grids}}\\
		\vspace{2mm}
		{\large Jack~Jackman$^{1}$, David~Ryan$^{1}$, Arun~Narayanan$^{2}$, Pedro~Nardelli$^{2}$, and~Indrakshi Dey$^{1}$}\\
		\vspace{7.5mm}
		$^1$\textit{Walton Institute, South East Technological University, Waterford, Ireland} \\
		$^2$\textit{Department of Electrical Engineering, Lappeenranta-Lahti University of Technology, Finland}
	\end{center}

\begin{abstract}
Modern power grids face an acute mismatch between where data is generated and where it can be processed: protection relays, EV (Electric Vehicle) charging, and distributed renewables demand millisecond analytics at the edge, while energy-hungry workloads often sit in distant clouds—leading to missed real-time deadlines and wasted power. We address this by proposing, to our knowledge, the first-ever SDEN (Software-Defined Energy Network) for CaaS (Computations-as-a-Service) that unifies edge, fog, and cloud compute with 5G URLLC (Ultra-Reliable Low-Latency Communications), SDN (Software-Defined Networking), and NFV (Network Functions Virtualization) to co-optimize energy, latency, and reliability end-to-end. Our contributions are threefold: (i) a joint task-offloading formulation that couples computation placement with network capacity under explicit URLLC constraints; (ii) a feasibility-preserving, lightweight greedy heuristic that scales while closely tracking optimal energy-latency trade-offs; and (iii) a tiered AI (Artificial Intelligence) pipeline—reactive at the edge, predictive in the fog, strategic in the cloud—featuring privacy-preserving, federated GNNs (Graph Neural Networks) for fault detection and microgrid coordination. Unlike prior edge-only or cloud-only schemes, SDEN turns fragmented grid compute into a single, programmable substrate that delivers dependable, energy-aware, real-time analytics—establishing a first-ever, software-defined path to practical, grid-scale CaaS. Based on simulation results, the hybrid SDEN achieves up to 69.65\% and median 30.2\% total energy savings versus cloud-only processing; integrating URLLC cuts transmission delay 15\% and jitter 25\%; SDN-enabled re-routing with multi-tier allocation boosts availability from 99.9\% (translates to 45 minutes) to ~48 $\mu$s downtime/year; and the federated GNN attains precision 0.96, recall 0.94, $F1$ 0.95, with 35 ms response—all while reducing backbone bandwidth via local preprocessing.
\end{abstract}
\textbf{keywords} - Smart grid, 5G, URLLC, SDEN, Grid optimisation

%

\section{Introduction}

Decarbonization of the smart energy grid will require several changes in the electricity infrastructure, including a rapid transition from conventional fossil-fuel powered plants to distributed energy sources (DERs) such as renewable energy sources (RES), battery energy storage systems (BESS), and power converters. Moreover, in many smart-grid paradigms, household consumers participate actively in the generation and management of energy. To successfully manage RES-based modern interactive smart grids, large amounts of data must be collected from the entire network and effectively coordinated \cite{Abir2021-dw}. However, this can lead to significant computing and data transmission bottlenecks, underscoring the need for developing intelligent systems to ensure the efficient use of ICT resources.

A commonly proposed solution to optimize the energy efficiency of smart grids without causing computing and transmission bottlenecks is to upgrade the existing electricity grid with high-speed communication networks, such as 5G and fibre optics, and IoT-enabled sensors and meters. To achieve this, several communications-network topologies have been proposed, including centralised, decentralised, and various hybrid architectures \cite{Guerrero2013-om}. 
In a centralised communications architecture, a distributed energy resource management system has a centralised view of the grid, enabling it to relatively easily coordinate and implement grid planning and scheduling. However, there are single points of failure, higher latencies and delays, and high chances of congestions or intermittent connection. In contrast, a decentralised architecture reduces latencies, delays, and chances of congestions and dropped connections, but at the cost of effective coordination \cite{Sabri2019-wv}.

A hybrid edge--fog--cloud architecture represents a tradeoff between the centralised and decentralised architectures \cite{laayati2025metaheuristic, boubaker2025comprehensive}. By placing local controllers on the ``edge'', within the range of a 5G base station, mobile communication can be enabled between them. Moreover, data is processed in proximity to where it is needed, thereby reducing latency for critical tasks such as anomaly mitigation, enhanced security and privacy. Further, longer-term scheduling and AI-driven predictive analytics can be performed in the cloud. However, this approach adds a high degree of complexity to the system and comes with increased infrastructure requirements. Enabling device-to-device communications and virtualisation features increases the utilisation of communications networks, resulting in greater energy draw from 5G base stations.

\begin{figure*}
    \centering
    \includegraphics[width=0.9\linewidth]{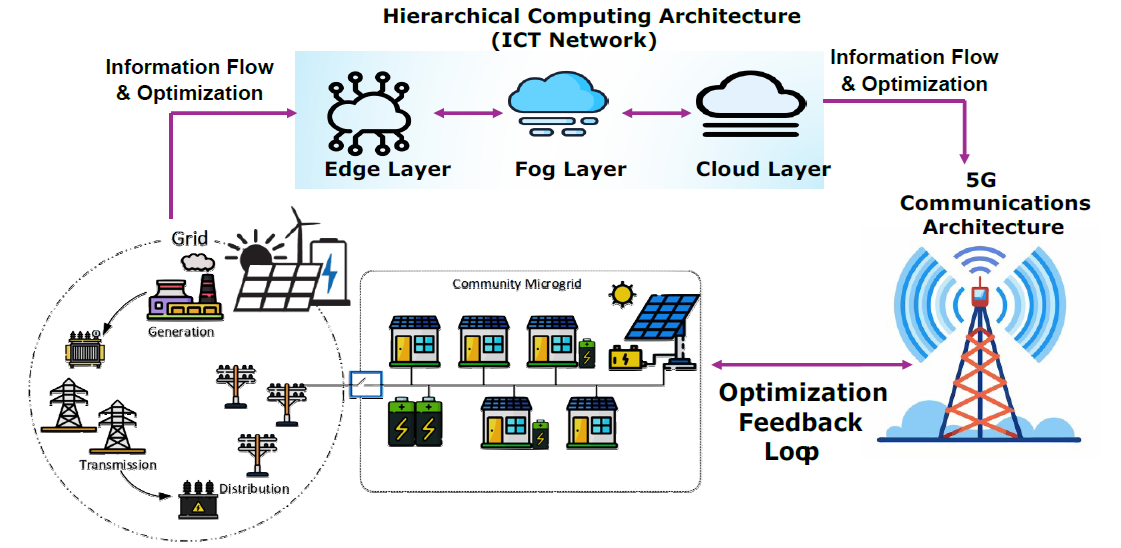}
    \vspace{-3mm}
    \caption{A Conceptual Representation of SDEN for CaaS}
     \vspace{-3mm}
    \label{fig:concept}
\end{figure*}

In this paper, we consider a software defined energy network (SDEN), an energy management concept inspired by software-defined networking in computer networks, with a hybrid edge--fog--cloud architecture. In SDENs, control, coordination, and optimization of energy flows (electricity, heat, renewables, storage, etc.) are handled by software-based control layers \cite{nardelli2021virtual}. We use \emph{Computations-as-a-Service (CaaS)} to mean the SDEN-controlled, SLA-driven placement and execution of grid applications, like, protection, automation, state estimation, analytics) across device, edge, and cloud resources—with the network fabric co-optimized to meet latency, reliability, and energy targets. Our hierarchical computing architecture comprises three layers---(1) an \textit{edge layer} that is reactive and predictive, comprising local devices such as DER-connected gateways (routers), sensors, actuators, etc.; (2) a \textit{fog layer} that comprises controllers such as microgrid-level controllers, to ensure key 5G features (ultra-low latency, ultra-reliability, etc.); and a (3) \textit{centralized cloud layer} that comprises data and control centres. These three layers are connected through data or control communication paths. Our proposed architecture (Fig. \ref{fig:concept}) will function to enable an energy and ICT-aware reactive smart grid system that will balance compute and communications resources through 5G virtualisation, have increased resilience to prevent outages, and support increased utilisation of RES.

Our SDEN is optimized to \textit{jointly} minimize total latency and energy across all edge-generated tasks, subjected to system and network constraints. We model this problem as an exact binary non-linear optimization problem with the objective to assign every task to the layer with the lowest combined energy-latency cost, a generally valid scenario in over-provisioned systems. The analytical solution set provides some critical insights on system design, but it is computationally expensive to scale it to larger networks. Hence, we also present a greedy heuristic algorithm that approximates the optimal solution in a scalable manner.

To validate the effectiveness of the proposed ICT--SG architecture, we employ the following four system performance metrics: (1) \textit{Energy Savings} that compares the energy usage with the hybrid edge-fog-cloud model versus the cloud-only baseline; (2) \textit{Active Generation Curtailment} that measures the percentage of available renewable energy that is curtailed (due to limitation in the ICT coordination or grid flexibility); (3) \textit{Packet Loss} that measures the packet data loss during transmission; and (4) \textit{Availability} that refines the uptime of a service or communication link.

Our primary contributions are as follows:
\begin{itemize}[leftmargin=*]
\item \textbf{Formalized, tiered distributed AI pipeline.} We formulate a mathematical and structural model for a distributed AI pipeline aligned with the hybrid mobile-edge ICT architecture. The pipeline is deployable across all three tiers—\emph{Edge} (reactive, real-time), \emph{Fog} (predictive, near-real-time), and \emph{Cloud} (strategic, batch/coordination)—with clearly delineated temporal and operational roles that integrate naturally with task offloading.

\item \textbf{Feasibility-preserving heuristic for joint energy--latency offloading.} We design and simulate a lightweight greedy heuristic that explicitly balances latency and energy when assigning tasks across the edge--fog--cloud tiers. Despite its simplicity, it consistently tracks an edge-first bandwidth optimum while delivering substantial energy reductions—\emph{up to} 69.65\%—under realistic capacity and deadline constraints.

\item \textbf{URLLC-integrated networking for real-time performance.} We provide the first simulation-based evidence that coupling ultra-reliable low-latency communications (URLLC) with intelligent offloading yields measurable network-layer gains: transmission delay decreases by 15\% and jitter by 25\%, strengthening support for hard real-time grid functions.

\item \textbf{Availability gains via SDN-enabled redundancy.} We establish, for the first time, that adding redundancy through software-defined networking (SDN) rerouting and multi-tier allocation elevates end-to-end availability from “three nines” (99.9\%) to effectively failure-proof levels, with annual downtime reduced to \emph{tens of microseconds} in our simulations.

\item \textbf{Federated GNNs for privacy-preserving grid intelligence.} We introduce and validate the first federated graph neural network (GNN) approach for fault detection and microgrid coordination in a hybrid SDEN, attaining state-of-the-art accuracy (precision 0.96, recall 0.94, F1 0.95) with \(\approx\)35\,ms response, while preserving data locality, privacy, and scalability.
\end{itemize}

Collectively, our results provide, to our knowledge, the first demonstration that a hybrid SDEN architecture, paired with lightweight optimization and distributed intelligence, can simultaneously address bandwidth overheads, energy inefficiency, latency sensitivity, and reliability constraints within a single, integrated framework.

The remainder of the paper is organized as follows. In Section \ref{sec:methodology}, we first develop our system formally with a novel binary non-linear optimization model that performs task placement, taking into consideration optimum energy–latency trade-off and bandwidth and URLLC feasibility. We also introduce a feasibility-preserving greedy heuristic with proven complexity bounds to derive approximate results for the optimum energy–latency trade-off in real time. Subsequently, we describe a tiered AI pipeline that applies the proposed hybrid edge-fog-cloud model to smart grid optimization and present two use cases---disaster-resilient microgrids and energy communities---in which we use distributed and federated GNNs for privacy-preserving fault detection and coordination.
In Section \ref{sec:results}, we describe our experimental setup, datasets, and the results obtained from applying our proposed methodologies and concepts; we validate our approach using our previously proposed baselines and metrics. Finally, Section \ref{sec:conclusions} concludes with a discussion of the limitations of the current study and the open challenges that motivate future work. 

\section{Related Work} \label{sec:related_work}


Numerous studies have investigated energy management and control in smart grids technologies as well as the integration of 5G networks to enable edge computing \cite{Shahgholian2021-xn} \cite {Chen2023-dd}. However, only a few studies have explored the possibilities to optimise the electrical grid and the ICT network using 5G functions like URLLC. {Beyond general surveys on smart-grid/edge computing, several works explicitly study co-optimizing grid functions and the ICT plane using 5G features—especially URLLC and slicing. Experimental and co-simulation studies show that IEC-61850 protection traffic (GOOSE/SV) can meet sub-10–20\,ms end-to-end targets over 5G (SA/NSA) when URLLC classes and resource engineering are applied, enabling teleprotection and substation automation under appropriate design assumptions \cite{Kazme2025-energies-5g-iec61850,Demidov2022-iec61850-5g-testbed}. Building on this, slice-aware resource control has been proposed and evaluated to prioritize protection over other utility traffic—e.g., hierarchical token bucket shaping and uplink bitrate adaptation inside a dedicated slice—to reduce delay/jitter and increase delivery ratio for GOOSE/SV frames \cite{Raussi2023-jeng-5g-slice-protection}. Wide-area designs further demonstrate R-GOOSE over cellular (5G/4G) with GRE/DMVPN, quantifying latency and packet-loss trade-offs and outlining security hooks \cite{Jafary2022-energies-rgoose-5g}. Analytical models of 5G RAN slicing with mixed numerologies capture priority for protection/control versus parallel traffic, providing design knobs that directly couple comms settings to power-system KPIs \cite{Mendis2021-arxiv-ran-slicing-smartgrid}. Complementary testbed studies and industry-facing evaluations report protocol and stack-level latencies for GOOSE, MMS, Modbus and DNP3 over 5G, informing scheduler and core-placement choices \cite{Boeding2023-osti-5g-protocol-latencies,Mendis2019-cired-5g-slicing-utility}. These results collectively motivate the co-design and co-optimization we pursue here.}

{An SDEN can jointly steer where computation runs and how traffic is carried to meet latency budgets while minimizing energy use. Moving analytics from cloud to edge/fog shortens control loops and reduces backhaul, a pattern established in edge/fog literature and recent power-system surveys \cite{bonomi2012fog,yildirim2025edgepower}. On the network side, SDN gives energy elasticity—turning off idle links/switches and consolidating flows—while preserving performance when engineered carefully (e.g., ElasticTree achieves up to \(\sim 50\%\) network-energy savings under real workloads) \cite{heller2010elastictree}. At the radio/edge layer, MEC resource-allocation surveys for 5G/6G document URLLC-aware mechanisms (e.g., slicing/QoS prioritization, configured-grant, short TTI) that reduce tail latency under power constraints \cite{sarah2023mecurllc}. Utility-specific studies go further: IEC-61850 R-GOOSE protection traffic has been demonstrated over 5G with quantified latency/packet-loss trade-offs \cite{jafary2022rgoose}, and slice-aware shaping/uplink-bitrate adaptation improves delivery ratio and latency for protection streams in shared networks \cite{raussi2023slice}. These findings motivate SDEN designs that co-optimize device–edge–cloud placement and transport/slice controls to jointly satisfy grid latency targets and energy objectives.}

Our study is related to resource management in edge--fog--cloud environments and the the need to optimize multiple, often conflicting objectives such as latency, energy consumption, cost, and security \cite{boubaker2025comprehensive}. This is complicated by the modern distributed computing systems, primarily due to the heterogeneous nature of underlying infrastructure and highly dynamic workloads. In \cite{laayati2025metaheuristic}, a smart Edge--Fog--Cloud energy management system that is tailored for energy-intensive industrial mining operations was designed, implemented, and validated. However, unlike our study, they did not jointly optimize energy and latency. Similarly, other works on task offloading in Edge--Cloud or Edge--Fog--Cloud architectures have focused on single-objective problems such as minimization of task completion time (\cite{lyu2022adaptive}), maximization of resource utilization (\cite{xiong2023multi}), and minimization of computational costs (\cite{li2022computing}). The authors in \cite{lin2024edge}, on the other hand, performed a multi-objective optimization in a edge--fog--cloud hybrid collaborative computing architecture that was applied to industrial  manufacturing services. They investigated computational task offloading with the aim to minimize latency and cost. In \cite{yang2020multi}, the authors optimized the total execution time and task resource cost in the fog network with a multi-objective task scheduling model. The optimal solution was obtained using a an improved multi-objective evolutionary heuristic algorithm and performance trade-offs were resolved using Pareto-based approaches. Similarly, multi-objective--- energy consumption, completion time, and economic cost---application placements was achieved with a novel Pareto-based approach in \cite{mehran2019mapo}, whereas the authors in \cite{chen2020deep} used a deep reinforcement learning-based dynamic resource management model to jointly optimize power control and computing resource allocations in mobile edge computing in industrial Internet of Things.

In contrast to the abovementioned studies, we present the first comprehensive exploration of an SDEN that integrates edge, fog, and cloud resources through an \textit{energy–latency–aware task offloading} heuristic for smart grid applications. We demonstrate that a fully hybrid architecture, jointly optimised with modern 5G features (URLLC, SDN, NFV) and distributed AI, can deliver simultaneous gains in latency, energy efficiency, bandwidth reduction, reliability, and fault detection.


\section{Methodology} \label{sec:methodology}

We first formalize the system model, comprising tiers (edge/fog/cloud), compute/communication resources, and 5G control primitives (URLLC, SDN, and NVF), as well as a task model that captures arrival processes, deadlines, and per-tier service times. We then derive formulations for energy and latency, considering computation and data transport, and describe a novel binary non-linear optimization that combines task placement with bandwidth and URLLC feasibility. Because exact solutions for binary integer non-linear programming do not scale, we introduce a feasibility-preserving greedy heuristic with proven complexity bounds that approximates the optimal energy–latency trade-off in real time. Finally, we describe a tiered AI pipeline, comprising a reactive edge, predictive fog, and a federating cloud, and the federated GNN that is used for privacy-preserving fault detection and coordination in smart grids.
\subsection{{Mathematical Formulation and Optimization Framework for Hybrid ICT-Enabled Smart Grid}}

{Let us model the hybrid ICT-enabled smart grid environment as a directed graph $\mathcal{G} = (\mathcal{N}, \mathcal{E})$, where $\mathcal{N}$ is the set of all computing nodes and $\mathcal{E}$ is the set of directed edges representing available data or control communication paths. $\mathcal{N}$ comprises three categories of devices: (1) edge devices, $\mathcal{N}_E$, such as DER-connected gateways, sensors, actuators, etc.; fog devices, $\mathcal{N}_F$, such as microgrid controllers; and central cloud or data centres $\mathcal{N}_C$. Now let an edge node $i \in \mathcal{N}_E$ generate tasks, such as measurements or control decisions, at a rate $\beta_i$. These tasks can either be processed locally or offloaded to fog or cloud nodes. Also, let us consider that within this environment, any type of node $n \in \mathcal{N}$ has a maximum computation capacity $\mu_n$, measured in tasks per second. Finally, let us define a task assignment variable, such that}
\[ {x_{i,n} = \begin{cases} 1, & \text{if task from node } i \text{ is assigned to node } n \\ 0, & \text{otherwise} \end{cases} }
\]
{$\text{for } i \in \mathcal{N}_E,\ n \in \mathcal{N}$. It is noteworthy that each task must be assigned to exactly one processing node (edge, fog, or cloud). The total latency $L_{i,n}$ for a task generated at edge node $i$ and executed at node $n$ can be expressed as follows:}

\begin{align}
{L_{i,n} \;=\; \frac{d_i}{b_{i,n}} \;+\; \frac{1}{\mu_n}}
\end{align}
{where $d_i$ is the data size per task (bits), $b_{i,n}$ is the available data rate (bits/s), and $\mu_n$ is the service rate at node $n$ (tasks/s).}
\begin{align}
{E_{i,n} \;=\; d_i\,\varepsilon^{\text{comm}}_{i,n} \;+\; c_i\,\varepsilon^{\text{proc}}_{n}}
\end{align}
{where $\varepsilon^{\text{comm}}_{i,n}$ is the energy per bit (J/bit) and $\varepsilon^{\text{proc}}_{n}$ is the energy per compute cycle (J/cycle); $c_i$ is the per-task compute demand (cycles).}

{Our main objective is to jointly minimize total latency and energy across all edge-generated tasks, subject to system and network constraints. Let us start by defining the decision variable, $x = \left\{x_{i,n}|i \in \mathcal{N}_E, n \in \mathcal{N}\right\}$. The objective function can then be expressed as, }
\begin{align}
    &{\min_x \sum_{i \in \mathcal{N}_E} \sum_{n \in \mathcal{N}} x_{i,n}\left(\omega_1L_{i,n} + \omega_2E_{i,n}\right)} \\
    {\text{subjected to}}&\nonumber\\
    &{\sum_{n \in \mathcal{N}} x_{i,n} = 1 \quad \forall i \in \mathcal{N}_E} \\
    &{\sum_{i \in \mathcal{N}_E} x_{i,n} \cdot \beta_i \leq \mu_n \quad \forall n \in \mathcal{N}} \\
    &{x_{i,n} \cdot L_{i,n} \leq L_{\text{max}} \quad \forall i, n }\\
    &{x_{i,n} \in \{0,1\} }
\end{align}

{Here, $\omega_1$ and $\omega_2$ are weights balancing latency and energy, respectively. 
Constraint (4) enforces exclusivity of assignment: each task stream $i$ is processed by exactly one node $n$. 
Constraint (5) enforces compute-capacity feasibility at each node by matching \emph{rates}:}
\begin{align}
{\sum_{i} x_{i,n}\,\beta_i \;\le\; \mu_n,\quad \forall n,}
\end{align}
{where $\beta_i$ is the arrival rate of tasks from edge node $i$ (tasks/s) and $\mu_n$ is the service rate of node $n$ (tasks/s). 
If a safety margin is desired, we impose a maximum limit, $\rho_{\max}$ on the utilization so that $0<\rho_{\max}\le 1$ and $\sum_i x_{i,n}\beta_i \le \rho_{\max}\mu_n$. 
Constraint (6) imposes latency bounds under URLLC: for any assignment $x_{i,n}=1$, $L_{i,n}\le L_{\max}$ (implemented, e.g., via a big-$M$ reformulation $L_{i,n}\le L_{\max}+M(1-x_{i,n})$). 
Constraint (7) imposes binary assignment $x_{i,n}\in\{0,1\}$. Here $\rho_{i, n} \in [0,1]$ denotes the offered-load/utilization (or an equivalent buffer-occupancy proxy) on the path from $i$ to $n$; in simplified experiments it may be set to a constant $\rho$.}

\subsubsection{{Analytical Solution}}

{We will first explore the analytical approach to solve the above hybrid task offloading optimization problem. The aim is to extract structural insights. In order to solve (3) tractably, we assume that (a) there are $N_E$ edge nodes indexed by $i \in \{1, \dotso, n\}$; (b) each node generates exactly one task, to be assigned to one of the three processing tiers: Edge (0), Fog (1), or Cloud (2); and (c) all the nodes within a tier are homogeneous in their processing and communication properties. We start by defining the decision variable,}
\[
{x_i = \begin{cases} 0, & \text{if task } i \text{ is assigned to Edge tier} \\ 1, & \text{if assigned to Fog tier} \\ 2, & \text{if assigned to Cloud tier} \end{cases}} 
\]
{
Let $L_i^k$ and $E_i^k$ denote the latency of and energy consumed by node $i$ on tier $k \in \{e,f,c\}$, respectively. Let us define a unified cost function for a task on a node-tier pair as follows: $\Phi_i^k = \omega_1L_i^k + \omega_2E_i^k$ where $\Phi_{i}^k$ denotes the cost function for the node $i$ on a tier $k$ to perform an assigned task, and $\omega_1$ and $\omega_2$ are weights reflecting the relative importance of latency and energy, respectively. Then the overall optimization problem can be redefined as follows:}
\begin{align}
    &\min_{t_i \in \{e,f,c\}} \sum_{i\in\mathcal N_E} \Phi_i(t_i), \nonumber\\
& {\Phi_i(k)=\Phi_i^k=\omega_1 L_i^k+\omega_2 E_i^k,\; k\in\{e,f,c\}.}
\end{align}
{subject to capacity constraints caused by the finite processing capacity of each tier. Here $i$ indexes edge-generated task streams ($\mathcal N_E$), while $k\in\{e,f,c\}$ indexes tiers; in the full per-node formulation we use binaries $x_{i,n}$ with $n\in\mathcal N=\mathcal N_E\cup\mathcal N_F\cup\mathcal N_C$, and the tier choice $t_i$ is the tier of the unique $n$ with $x_{i,n}=1$.}
\begin{align}
    {\sum_{i \in \mathcal{N}_E} 1(t_i = e) \leq C^e, \sum_{i \in \mathcal{N}_E} 1(t_i = f) \leq C^f, \sum_{i \in \mathcal{N}_E} 1(t_i = c) \leq C^c} 
\end{align}

{where $C^E$, $C^f$, and $C^c$ are the maximum number of tasks that can be served by Edge, Fog, and Cloud, respectively. To handle the capacity constraints, we introduce Lagrangian multipliers, $\lambda_e$, $\lambda_f$, $\lambda_c$ for the Edge, Fog, and Cloud respectively. The Lagrangian function then can be expressed as follows:}
\begin{align}
    {\mathcal{L}(x, \lambda) = \sum_{i \in \mathcal{N}_E} \Phi_i(x_i)} &{+ \lambda_e \left( \sum_{i \in \mathcal{N}_E} 1(t_i = e) - C^e \right)} \\ &+ {\lambda_f \left( \sum_{i \in \mathcal{N}_E} 1(t_i = f) - C^f \right)} \\ &{+ \lambda_c \left( \sum_{i \in \mathcal{N}_E} 1(t_i = c) - C^c \right)} 
\end{align}
{Rewriting, we group terms by node to obtain,}
\begin{align}
    {\mathcal{L}} &{= \sum_{i \in \mathcal{N}_E} \bigg[ \Phi_i^e 1(t_i = e) + \Phi_i^f 1(t_i = f) + \Phi_i^c 1(t_i = c)} \nonumber\\
    &{+ \lambda_e 1(t_i = e) + \lambda_f 1(t_i = f) + \lambda_c 1(t_i = c) \bigg]} \nonumber\\
    &{- \sum_{k \in \{e, f, c\}} \lambda_k C^k} 
\end{align}
{where $C^k$ denotes the tasks served by the $k$th tier and (12) simplifies to,}
\begin{align}
    {\mathcal{L} = \sum_{i \in \mathcal{N}_E} \min_{k \in \{e,f,c\}} \left( \Phi_i^k + \lambda_k \right) - \left( \lambda_e C^e + \lambda_f C^f + \lambda_c C^c \right)} 
\end{align}
{Now, let us define the dual function,} 
\begin{align}
    {g(\lambda) = \sum_{i \in \mathcal{N}_E} \min_{k \in \{e,f,c\}} \left( \Phi_i^k + \lambda_k \right) - \sum_{k \in \{e, f, c\}} \lambda_k C^k} 
\end{align}

{We solve the dual maximization $\max_{\lambda\ge 0}\, g(\lambda)$ with $\lambda=(\lambda_e,\lambda_f,\lambda_c)$.
Given tier choice $t_i\in\{e,f,c\}$ for each edge task $i\in\mathcal N_E$, the dual function is $g(\lambda)=\sum_{i\in\mathcal N_E}\min_{k\in\{e,f,c\}}\bigl(\Phi_i^k+\lambda_k\bigr)\;-\;\sum_{k\in\{e,f,c\}}\lambda_k\,C_k,$, where $C_k$ is the (aggregate) capacity of tier $k$ measured as the maximum number of tasks assignable at that tier. The sub-gradient of $g$ is $s_k(\lambda)=\sum_{i\in\mathcal N_E}\mathbf{1}\!\left(t_i^*(\lambda)=k\right)-C_k,
\quad t_i^*(\lambda)=\arg\min_{k\in\{e,f,c\}}\bigl(\Phi_i^k+\lambda_k\bigr).$}

{We update multipliers by projected sub-gradient $\lambda_k^{(j+1)}=\Bigl[\lambda_k^{(j)}+\sigma_j\,s_k\bigl(\lambda^{(j)}\bigr)\Bigr]_+,
\quad k\in\{e,f,c\},$ with step sizes $\sigma_j>0$ (e.g., diminishing, $\sigma_j\downarrow 0$, $\sum_j\sigma_j=\infty$). We stop when $\|s(\lambda^{(j)})\|_\infty\le \varepsilon$ (or after a preset iteration budget). The primal assignment is then recovered by $t_i^\star=\arg\min_{k\in\{e,f,c\}}\bigl(\Phi_i^k+\lambda_k^\star\bigr),\quad i\in\mathcal N_E.$ If any tier remains over capacity (rare in practice due to projection), we apply a lightweight repair: for each overfull tier $k$, reassign the smallest number of tasks with the largest \emph{cost gaps} 
$\Delta_i(k\!\to\!k')=\bigl(\Phi_i^{k'}+\lambda_{k'}^\star\bigr)-\bigl(\Phi_i^{k}+\lambda_{k}^\star\bigr)$ to the next-best feasible tier $k'$ until $\sum_i\mathbf{1}(t_i^\star=k)\le C_k$. In over-provisioned settings where $C_e+C_f+C_c\ge |\mathcal N_E|$ and tier budgets are nonbinding, the optimal solution reduces to per-task selection
$t_i^\star=\arg\min_{k\in\{e,f,c\}}\Phi_i^k$, assigning each task to the minimum latency–energy tier. The multipliers $\lambda_k$ quantify the shadow price of capacity at each tier: if, e.g., edge is latency/energy-favored but scarce, a larger $\lambda_e$ discourages overuse by shifting marginal tasks to fog or cloud.}

{In the analytical model, we assume one task stream per edge node, so that the total number of tasks is $N_E$. Feasibility requires that the aggregate capacity across all tiers can accommodate all tasks, i.e. $N_E \le C^e + C^f + C^c$. 
When this inequality holds strictly, the system is over-provisioned, and capacities do not bind; the optimal assignment then simplifies to choosing the lowest-cost tier per task. When the inequality holds with equality, capacities are exactly saturated. If $N_E > C^e + C^f + C^c$, the system is under-provisioned: not all tasks can be scheduled. In such cases, our current formulation would treat the problem as infeasible; in practice, this would correspond to dropping or deferring some tasks, or prioritizing a subset (e.g., URLLC flows) while rejecting others. 
For clarity, our analytical solution focuses on the feasible/over-provisioned regimes, while under-provisioning scenarios are better addressed through admission control or priority-based extensions of the model.}

\subsubsection{Heuristic Solution Strategy}

{As seen in the above subsection, due to the binary and non-linear nature of the optimization problem, solving it exactly is computationally intensive. Therefore, next, we develop a greedy heuristic algorithm that approximates the optimal solution in a scalable manner.}

\begin{algorithm}[h]
\caption{~~{Energy-Latency Trade-off}}\label{tab:x}
{\textbf{Input} : Task generation rates $\beta_i$, node capacities $\mu_n$, data transfer rates $b_{i,n}$ and energy metric $\eta$.}\\
{\textbf{Step 1} : For each edge node $i$: Evaluate all $n \in \mathcal{N}$ that satisfy $L_{i,n} \leq L_{\text{max}}$ and $\beta_i \leq \mu_n$.}\\ 
{\textbf{Step 2} : Compute cost metric : $\Phi_{i,n} = \omega_1 L_{i,n} + \omega_2 E_{i,n}.$}\\
{Select $n^* = \arg\min_n \Phi_{i,n}$}\\
{Assign $x_{i,n^*} = 1$, update $\mu_{n^*} \leftarrow \mu_{n^*} - \beta_i$} \\
{\textbf{Step 3} : Return final assignment $x$.}
\end{algorithm}

{Note that the optimization does not enforce edge-first assignment. Edge is chosen only when it minimizes the combined latency–energy cost \(\Phi_i^e\) \emph{and} respects capacity. When edge resources are scarce or relatively expensive, the formulation naturally diverts some tasks to fog or cloud. Thus, edge is often—but not always—the selected tier, consistent with a balanced capacity–cost design.}

\subsection{{Performance Metrics for Grid Optimization}}

{To validate the effectiveness of the hybrid mobile edge ICT architecture introduced in this work, we present formal mathematical definitions of system performance metrics. We define \textbf{Energy Savings} as follows:}
\begin{align}
{\Delta E = E^{\text{base}} - E^{\text{opt}} \quad [\text{kWh}] }
\end{align}
{where $E^{\text{base}}$ is the energy consumption using centralized cloud-based processing and $E^{\text{opt}}$ is the energy consumption using hybrid ICT architecture. For $N$ tasks with arrival rates $\beta_i$, we define,}
\begin{align}
   { E^{\text{base}}} &{= \sum_{i=1}^{N} \left( \beta_i \cdot \eta_{i,\text{cloud}}^{\text{comm}} + \frac{\beta_i}{\mu_c} \cdot \eta_{\text{cloud}}^{\text{proc}} \right) }\nonumber\\
    {E^{\text{opt}}} &{= \sum_{i=1}^{N} \left( \beta_i \cdot \eta_{i,n(i)}^{\text{comm}} + \frac{\beta_i}{\mu_{n(i)}} \cdot \eta_{n(i)}^{\text{proc}} \right)} 
\end{align}
{where $n(i) \in \{edge, fog, cloud\}$ is the optimal destination node for task $i$, $\mu_n$ is the processing capacity of node $n$, and $\eta_{i,n(i)}^{\text{comm}}$, $\eta_{n(i)}^{\text{proc}}$ are the energy per bit for communication and per cycle for computation. The main aim of this metric is to evaluate the total reduction in energy usage when adopting the hybrid edge-fog-cloud model versus cloud-only baseline. }

{We define the \textbf{Active Generation Curtailment} as follows:}
\begin{align}
    {\text{Curtailment (\%)} = \left( \frac{P_{\text{gen}} - P_{\text{exported}}}{P_{\text{gen}}} \right) \times 100 }
\end{align}
{where $P_{\text{gen}}$ is the potential power generation by DERs and $P_{\text{exported}}$ is the power successfully injected into the grid. This metric measures the percentage of available renewable energy that is curtailed due to limitations in the ICT coordination or grid flexibility. If the ICT system introduces latency, $L_i > L_{\text{max}}$, fault handling and demand response operations may fail, triggering curtailment; $L_i > L_{\text{max}} \Rightarrow \text{Curtailment Event}$. This can result from congestion, poor bandwidth, or under-provisioned edge resources.}

{Additionally, we define \textbf{Packet Loss} as follows:}
\begin{align}
    {\text{Packet Loss (\%)} = \left( \frac{\sum_{i=1}^{N} \beta_i \cdot p_{i,n}}{\sum_{i=1}^{N} \beta_i} \right) \times 100} 
\end{align}
{where $p_{i,n}$ is the drop probability from node $i$ to $n$ and $\rho_{i,n}$ is the traffic load or buffer occupancy that is used to formulate $p_{i,n} = 1 - \exp(-\rho_{i,n})$ based on a common queuing-based model, where $p_{i,n}$ increases with network congestion.}

{Also, we define \textbf{Availability (\%)} which reflects the uptime of a service or communication link to define}
\begin{align}
    {\text{Availability (\%)} = \left( \frac{T_{\text{up}}}{T} \right) \times 100} 
\end{align}
{where $T_{\text{up}}$ is the uptime and alternatively,}
\begin{align}
    {\text{Availability} = \frac{\text{MTBF}}{\text{MTBF} + \text{MTTR}} \times 100 }
\end{align}
{where $\text{MTBF}$ is the meantime between failures and $\text{MTTR}$ is the meantime to repair. }

\subsection {{Distributed AI Pipeline for Grid Optimization}}

{
We now formulate a mathematical and structural model for a distributed AI pipeline for the hierarchical computing structure of the hybrid mobile edge ICT architecture. This model is designed for a specific Smart Grid requirement---intelligent grid monitoring, fault prediction, and control. Our proposed distributed AI pipeline can be deployed across the three architectural tiers---Edge, Fog and Cloud---with each tier being responsible for specific temporal and operational role, as listed in  Table~\ref{tab:1}.}
\begin{table}[h!]
\centering
\caption{{AI Roles Across Hierarchical Tiers}}
\begin{tabular}{@{}llll@{}}
\toprule
\textbf{Tier} & \textbf{AI Role} & \textbf{Functional Scope} & \textbf{Characteristics} \\
\midrule
Edge AI & Reactive & Fault detection & Low latency, local models \\
Fog AI & Predictive & Forecasting, monitoring & Moderate compute/latency \\
Cloud AI & Strategic & Optimization, control & High compute, tolerant \\
\bottomrule
\end{tabular}
\label{tab:1}
\end{table}

{Let us consider the same system model where $\mathcal{N}_E$, $\mathcal{N}_F$, $\mathcal{N}_C$ represent sets of edge, fog, and cloud nodes respectively. The distributed AI system aims to minimize operational risks and resource consumption by optimizing (a) fault detection and response time, (b) predictive maintenance scheduling, and (c) actions for controlling the grid and computing its resource allocations. Let $z_i(t)$ be the time-series input from sensor $i$ at time $t$ and $z_i(t+\tau)$ be the predicted state at time $(t+\tau)$. Let us start with the case of reactive detection and prediction at the Edge layer. Each DER or grid sensor node $i \in \mathcal{N}_E$ executes a local, low-complexity model: $y_i(t) = \mathrm{f}^e(x_i(t), \theta_E)$, where $y_i(t) \in \{0,1\}$ is a binary variable indicating detected fault or anomaly, $\mathrm{f}^e$ is the lightweight classifier, and $\theta_E$ are the model parameters optimized for speed and efficiency. If $y_i(t) = 1$, DER will be isolated or local protection protocol will be initiated with a targeted latency of $L_E < 10$ ms.}

{Next. we look at fog nodes that collect aggregate data from a local group of edge nodes $\{i_1, i_2, \dotso, i_m\} \subset \mathcal{N}_E$ resulting in $z_j(t) = [z_{i_1}(t), z_{i_2}(t), \dotso, z_{i_m}(t)]$. In this scenario, the fog model will be able to predict the fault likelihood at a future time instance of $\tau$, $\hat{y}_j(t+\tau) = \mathrm{f}^f(z_j(t:t - T); \theta_F)$, where $\mathrm{f}^f$ is the temporal learning model (different types of learning algorithms can be implemented, as shown in Section \ref{sec:results}); $\hat{y}_j \in [0,1]$ is the probability of fault occurrence; and $\theta_F$ are the parameters trained using historical and streaming edge data. If $\hat{y}_j(t+\tau) > \varepsilon$, then preventive maintenance or topology configuration is triggered. The goal is to achieve a latency of $L_F$ in the range of 100-300 ms with forecast horizon $\tau >> L_F$.}

{At the cloud level, let us consider that the system aggregates regional observations given by $Z(t) = \bigcup_{j \in \mathcal{N}_F} z_j(t)$. The cloud model learns long-range patterns for decision support $\hat{Y}(t+\Delta) = \mathrm{f}^c(X(t:t-T'); \theta_C)$ where $\mathrm{f}^c$ represents a Graph Neural Network (GNN) or transformer-based architectures (we compare the algorithms in Section \ref{sec:results}); $\hat{Y}$ is the output vector of grid-level KPIs, such as load forecast and voltage stability; and $\theta_C$ is the high-dimensional set of model parameters from training. Based on the learning of the long-range patterns, it will be possible to take wide-impact actions like grid-wide reconfiguration, load balancing or demand response and updating fog/edge model weights via hierarchical feedback.}

{With the above scenarios in place, we formulate the pipeline optimization problem. Let us define a composite cost function, $C_F$, over a prediction horizon $T$, balancing the cost of the fault or anomaly; a cost of the maintenance operation, $C_M$; and the cost of penalty from misprediction error, $C_E$. In this case, the objective function becomes}
\begin{align}
    {\min_{\theta_E, \theta_F, \theta_C} \mathbb{E} \Big[\sum_{t=1}^T C_F (y(t), \hat{y}(t)) + C_M(\alpha(t)) + C_E (y(t), \hat{y}(t))\Big]}
\end{align}
{subjected to the following latency constraints:}
\begin{align}
    \label{eq:latency-sum}
  L_{i,n}
   = 
  \underbrace{\frac{S_i}{R_{i,n}}}_{\text{transmission time (s)}}
   + 
  \underbrace{\frac{C_i}{\nu_n}}_{\text{service time (s)}}
   \le L_k^{\max},
   \forall i,\; n\in\mathcal{N}_k
\end{align}
{where $S_i$ [bits] is the payload size (or burst size) of task/flow $i$; 
$R_{i,n}$ [bits/s] is the achievable link rate available to $i$ at node/link $n$ (accounting for scheduling/slicing); 
$C_i$ [CPU-cycles] is the compute demand of $i$; 
$\nu_n$ [CPU-cycles/s] is the effective service rate at node $n$; 
$L_{i,n}$ [s] is the resulting end-to-end latency; 
$L_k^{\max}$ [s] is the latency budget for class $k$; 
$\mathcal{N}_k$ is the set of candidate nodes admissible for class $k$. Consequently, the following compute constraint (which defines that the total assigned load must not exceed each node's computing capacity):}
\begin{align}
    {\sum_{i \in \mathcal{N}_E} z_{i,n} \cdot \gamma_i \leq \mu_n, \quad \forall n \in \mathcal{N}_k}
\end{align}
{where $z_{i,n} \in \{0, 1\}$ is the task assignment indicator; $\mathcal{N}_k$, the set of nodes in tier $k \in \{e, f, c\}$; $\gamma_i$, the task load (data rate) of task $i$; $b_{i,n}$, the available data transfer rate between task origin $i$ and node $n \in \mathcal{N}_k$; $\mu_n$, the computing capacity of node $n$; and $L_{i,n}$, the end-to-end latency for task $i$ assigned to node $n$. The accuracy constraint is given by}
\begin{align}
    {\mathbb{P}(\text{True Positive (TP)}) = \frac{\sum_{i} \mathbf{1}[y_i = 1 \land \hat{y}_i = 1]}{\sum_{i} \mathbf{1}[y_i = 1]}}
\end{align}
{where $y_i(t) \in \{0,1\}$ is the true fault label for node $i$ at time $t$, $\hat{y}_i(t) \in \{0,1\}$ is the predicted label from AI model, TP is the number of true positives (Correctly detected faults), and $\mathbb{P}$ is the total number of actual positive cases. Based on the above, we can enforce $\mathbb{P}(\text{True Positive}) \geq \delta$, where $\delta \in [0,1]$ is the minimum acceptable true positive rate (e.g., $\delta = 0.95$).}

{In the proposed edge-fog-cloud architecture, a continuous feedback and adaptation mechanism is established across computational layers to ensure dynamic model refinement and enhanced grid resilience. Specifically, when an event or anomaly is detected, the edge layer initiates a localized alert, which is subsequently validated and contextualized at the fog layer through regional aggregation and short-term predictive modelling. The fog layer, upon forecasting anomalies or identifying evolving patterns, transmits these insights to the cloud layer, where comprehensive retraining of global models occurs. Upload model parameters $\theta$ are then disseminated downstream to refresh the operational intelligence at both the fog and edge layers.}

{The model update schemes are formally defined as follows. At the edge layer, models are fine-tuned in situ via an on-device gradient descent update:}
\begin{align}
    {\theta_E (t+1) \gets \theta_E(t) - \omega\nabla_{\theta_E}\chi_E}
\end{align}
{where $\omega$ represents the learning rate and denotes the local loss function $\chi_E$ at the device level. At the fog layer, parameters are adapted through streaming mini-batch learning by solving}
\begin{align}
    {\theta_F \gets \text{argmin}_{\theta} \sum_{j=1}^\mathfrak{B} \chi_F(\hat{y}_j, y_j)}
\end{align}
{where $\mathfrak{B}$ is the mini-batch size, $y_j$ are the predicted outputs, and $y_j$ are the corresponding actual labels. At the cloud layer, model retraining occurs either through conventional batch offline learning or via federated aggregation, wherein parameter updates from multiple fog and edge nodes are consolidated without necessitating raw data sharing. This hierarchical, closed-loop adaptation framework ensures that the Smart Grid environment remains responsive, context aware, and capable of continuous learning in highly dynamic and uncertain operating environments.}

\subsection{{Complementarity of Predictive Fog AI and Reactive Edge AI for Risk-aware Grid Management}}

{The operational resilience of smart grid is contingent not only upon the speed of local fault detection mechanisms but also critically dependent on predictive capabilities that anticipate system disturbances. This subsection rigorously formulates how predictive AI, deployed at the fog layer, and reactive AI, operating at the edge layer, act as complementary agents to minimize operational risk and maximize system reliability within the hybrid ICT infrastructure. To this end, we distinguish the two AI layers according to their temporal response characteristics, computational complexity, and decision-making objectives. Specifically, the edge layer provides immediate fault detection and protection actions with reaction on the order of milliseconds, whereas the fog layer delivers short-term forecasts of potential faults or instabilities over seconds to minutes. Although each layer operates autonomously, their coordinated interaction establishes a distributed cognitive feedback loop that significantly enhances overall grid intelligence and responsiveness.}

{Formally, let $w_i(t)$ denote the sensor measurement (e.g., voltage, frequency) collected by DER node $i$ at time $t$. The true operational state is indicated by $v_i(t) \in \{0,1\}$ where $v_i(t) = 1$ signifies an anomaly or fault condition. The edge AI model generates a binary anomaly detection output;}
\begin{align}
    {\hat{v}^E_i(t) = \psi_E(w_i(t); \phi_E) \in \{0, 1\}}
\end{align}
{where $\psi_E(\cdot)$ denotes a lightweight binary classification function, parameterized by $\phi_E$. Upon detection of an anomaly $(\hat{v}^E_i(t) = 1)$, a proactive action such as microgrid isolation or DER disconnection takes place, ensuring a reaction time of less than 10 ms to satisfy URLLC constraints.}

{In parallel, fog-level AI models aggregate time-series observations across spatially proximate nodes defining the aggregated input, $w_j(t) = \{w_i(t-k)\}_{k=0}^T \forall i \in \text{cluster}_j$ and forecasting the future likelihood of anomalies as, }
\begin{align}
    {\hat{v}^F_i(t + \tau) = \psi_F(w_j(t:T); \phi_F), \hat{v}^F_j \in [0, 1]}
\end{align}
{where $\psi_F(\cdot)$ denotes a recurrent neural network with learned parameters $\phi_F$. A predicted probability of $\hat{v}^F_i(t + \tau)$ exceeding a threshold $\varepsilon$ triggers preventive control actions including dynamic reconfiguration, predictive maintenance dispatch, or adaptive model updates at the edge.}

{The risk trade-offs between these AI layers are captured through a probabilistic cost model. Let $\mathcal{P}_F$ denote the probability of a fault occurring within the prediction window $[t,t+\tau]$, $\mathcal{P}_E$ the probability of successful real-time fault detection at the edge, and $\mathcal{P}_F^{\text{pred}}$, the probability of successful fault prediction at the fog layer. Furthermore, let us define the following cost parameters: $\Omega_{\text{undetected}}$ is the cost of missed detection; $\Omega_{\text{proactive}}$, the cost of false alarm and unnecessary proactive action; and $\Omega_{\text{reactive}}$, the cost of mitigation after detection. The expected costs under various operational strategies are formalized as follows:}
\begin{enumerate}
    \item \textbf{Edge AI Only:} $\Omega_{\text{edge}} = \mathcal{P}_F \cdot (1 - \mathcal{P}_E) \cdot \Omega_{\text{undetected}} + \mathcal{P}_E \cdot \Omega_{\text{reactive}}$
    \item \textbf{Fog AI Only:} $\Omega_{\text{fog}} = \mathcal{P}_{F_{\text{pred}}} \cdot \Omega_{\text{proactive}} + (1 - \mathcal{P}_{F_{\text{pred}}}) \cdot \mathcal{P}_F \cdot \Omega_{\text{undetected}}$
    \item \textbf{Combined Fog + Edge AI:} $\Omega_{\text{combined}} = \mathcal{P}_{F_{\text{pred}}} \cdot \Omega_{\text{proactive}} + (1 - \mathcal{P}_{F_{\text{pred}}}) \cdot \mathcal{P}_F \cdot (1 - \mathcal{P}_E) \cdot \Omega_{\text{undetected}} + \mathcal{P}_E \cdot \Omega_{\text{reactive}}$
\end{enumerate}
{Visual evidence demonstrates that $\Omega_{\text{combined}} \leq \min(\Omega_{\text{edge}}, \Omega_{\text{fog}})$ which confirms that the integration of predictive and reactive AI layers systematically reduces the expected operational risk and associated mitigation cost.}

{Moreover, the reliability of system fault management can be quantified as a function of the detection success probability as follows: $\varphi_e (t) = \mathcal{P}_E and \varphi_f (t) = \mathcal{P}_F^{\text{pred}}$ for the edge-only and fog-only cases, respectively, while the reliability under combined operation is given by $\varphi_{\text{combined}}(t) = 1 - (1 - \mathcal{P}_E)(1 - \mathcal{P}_F^{\text{pred}})$, which guarantees that $\varphi_{\text{combined}}(t) \geq \max(\varphi_e (t), \varphi_f (t))$. Thus, the layered AI framework not only distributes computational and decision-making burdens across the edge-fog-cloud hierarchy but also increases overall grid resilience through synergistic intelligence. This layered approach is fundamental to enabling self-healing, resilient smart grid ecosystem under dynamic and distributed energy landscapes.}\vspace{-4mm}

\subsection{Use-cases Enabled by AI-driven Optimization}

{Building upon the secure, intelligent, and scalable ICT architectures developed throughout this work, we propose two advanced use cases that exemplify the hybrid edge-fog-cloud architecture's transformative potential for smart grids. These futuristic scenarios involve dynamic and distributed systems demanding high resilience, autonomy, and real-time topology-aware optimization. To meet these requirements, we plan to use GNN as a distributed, scalable inference mechanism across all ICT layers. We present two representative scenarios: Disaster-Resilient Microgrids and AI-Optimized Energy Communities, each formally modelled as optimization formulations and GNN-based decision-making frameworks.}

\subsubsection{Disaster-Resilient Microgrids}

{In this scenario, microgrids operating under extreme conditions, such as natural disasters, are modelled as graphs, $\hat{\mathcal{G}} = (\hat{\mathcal{N}}, \hat{\mathcal{E}})$, where $\hat{\mathcal{N}}$ denotes the set of DERs, loads, and relay, and $\hat{\mathcal{E}}$ captures physical and communication links. Each node $u \in \hat{\mathcal{N}}$ is characterized by generation $R_u^{\text{gen}}$, load $R_u^{\text{load}}$, and phase angle $\vartheta_u$ with power flow along the edge ($u, v$) denoted as $\nu_{u, \varpi}$. The microgrid's islanded operation is formulated as the constrained optimization problem:}
\begin{align}
    &{\min_{\nu_{u, \varpi}, R_u^{\text{shed}}}\sum_{u \in\mathcal{N}} \Big(R_u^{\text{shed}} + \kappa_u|\nu_{u, \varpi}|\Big)}\\
    \text{subject to}~&~{R_u^{\text{gen}} - R_u^{\text{shed}} = R_u^{\text{load}} + \sum_{\varpi \in \mathcal{N}} A_{u, \varpi}\nu_{u, \varpi} \forall u \in \mathcal{N}}\nonumber\\
    &~{|\nu_{u, \varpi}| \leq \nu_{u, \varpi}^{\text{max}}, \nu_{u, \varpi} = D_{u, \varpi}(\vartheta_u - \vartheta_{varpi})}\nonumber
\end{align}
{where $R_u^{\text{shed}} \geq 0$ is the amount of load shed at node $u$, $\kappa_u \geq 0$ is a weight representing the power loss over lines, and $A_{u, \varpi} \in \{0,1\}$ is the adjacency matrix entry where $A_{u, \varpi} = 1$ if nodes $u$ and $\varpi$ are connected and $D_{u, \varpi}$ is the line susceptance. To solve this problem in a distributed manner, each node embeds local features $h_u^{\text{(0)}} = [R_u^{\text{gen}}, R_u^{\text{load}}, \text{status}_u]$ and performs GNN-based message passing:}
\begin{align}
    {h_u^{(l+1)} = \sigma \bigg(\sum_{\varpi \in \mathcal{N}(u)} \mathcal{W}^{(l)} h_u^{(l)} + d_u^{(l)}\bigg)}
\end{align}
{where $\mathcal{W}^{(l)}$ and $d_u^{(l)}$ are learnable weights and biases, and $\sigma(\cdot)$ is a non-linear activation function at the $l$th layer. After the $\mathbb{L}$th layer, a multilayer perceptron (MLP) outputs flow and shedding decisions : $(\nu_{u, \varpi}, \hat{R}_u^{\text{shed}}) = MLP(h_u^{(\mathbb{L})})$, which is a distributed GNN-based solution enabling fast, autonomous microgrid reconfigurations that minimize reliance on cloud connectivity.}

\subsubsection{AI-Optimized Energy Communities with Federated GNNs}

{With the rise of prosumers, households that produce and consume energy, there is a need for distributed coordination of storage, trading, and consumption scheduling, respecting privacy, minimizing latency, and supporting millions of nodes. Let $Q = \{1, \dotso, N\}$ denote the set of prosumer nodes. Each node $u$ optimizes the energy imported from the main grid at time $t$, $R_u^{\text{import}}(t)$;  the energy exported to the main grid at time $t$,  $R_u^{\text{export}}(t)$; the storage action positive for charging and negative for discharging, $S_u(t)$; and consumption load, $\rho_u(t)$. The objective is to minimize total system cost:}
\begin{align}
    &{\min{\sum_{t=1}^T}\sum_{u \in Q} \bigg(q_t^{\text{grid}}R_u^{\text{import}}(t) - \bar{r}_tR_u^{\text{export}}(t)} \nonumber\\
    &\quad {+ \xi\cdot\text{Deviations}(\rho_u(t))\bigg)}
\end{align}
subject to
\begin{align}
    &{R_u^{\text{gen}}(t) + R_u^{\text{import}}(t) + S_u(t) = \rho_u(t) + R_u^{\text{export}}(t)} {\text{SOC}_u(t+1) = \text{SOC}_u(t) + \bar{\eta}_{\text{ch}}S_u^{\text{ch}}(t) - 1/\bar{\eta}_{\text{dis}}S_u^{\text{dis}}(t)}
\end{align}
{where $q_t^{\text{grid}}$ is the grid energy price; $\bar{r}_t$ is the reward (e.g., feed-in tariff) for energy exports; $\xi$ is a penalty coefficient for deviations from scheduled loads, $\text{SOC}_u(t)$ is the battery state of charge; and $\bar{\eta}_{\text{ch}}$ and $\bar{\eta}_{\text{dis}}$ are charging and discharging efficiencies. Each node $u$ initializes its feature vector $h_u^{\text{(0)}} = [R_u^{\text{gen}}, \text{SOC}_u(t), \hat{\rho}_u(t), t]$, where $\hat{\rho}_u(t)$ is the forecasted load. Through GNN layers, we can do message passing using (31) and final inference given by $\big(\hat{R}_u^{\text{import}}, \hat{S}_u(t), \hat{\rho}_u(t)\big) = MLP(h_u^{(\mathbb{L})})$. So, in this case, each household performs inference locally at the edge, while fog nodes periodically aggregate model weights without accessing raw data ensuring privacy.}

{Having established the system model, optimization formulation, and feasibility-preserving heuristic—along with the tiered AI pipeline and federated GNN, we will now evaluate the methodologies and present the Results.}

\section{Results and Discussion}\label{sec:results}

This section provides the results of applying the methodology detailed in Section \ref{sec:methodology} in a Python-based simulation, comparing baseline data with scenarios that apply the proposed optimisations. Here, we detail the experimental setup (workloads, network topologies, and parameter ranges), enumerate baselines (edge-only, cloud-only, and bandwidth-optimal placements), and present quantitative outcomes on end-to-end latency, energy consumption, jitter, availability, and fault-detection accuracy. We also analyze sensitivity to URLLC constraints and load spikes, and provide ablation studies that isolate each architectural component’s contribution to overall performance.

The results are divided into four subsections: (i) first, we demonstrate how the hybrid edge-fog-cloud architecture and task offloading algorithm improve system-wide performance in terms of energy, latency, and resource utilisation; (ii) then, we show how 5G features and intelligent networking improve resilience, packet delivery, and grid responsiveness; (iii) then, we evaluate the performance of the distributed AI pipeline, showing how combining edge and fog AI improves fault management; and (iv) finally, we showcase the effectiveness of federated GNNs in coordinating distributed energy communities.

\subsection{Simulation Setup}

To rigorously evaluate the proposed hybrid edge–fog–cloud architecture, we design a simulation environment that replicates realistic task generation, communication, and processing in a smart grid context. The setup is structured to capture tradeoffs between latency, energy consumption, and resource availability, while enabling comparison across baseline and optimised scenarios. Below, we describe in detail the chosen parameters, their rationale, and the experimental methodology.

\subsubsection{Task Generation Model}
Tasks are generated at the edge layer in the form of measurements, control signals, and anomaly detection events produced by DERs and IoT sensors. The inter-arrival rate of tasks is set between one every 0.5 and 1.5 s ($\beta_i \in [0.67, 2]$ tasks/second). This stochastic range reflects the heterogeneity of data generation in practice: frequent updates are needed for real-time fault detection and load balancing, while slower rates capture routine monitoring data. The range is narrow enough to create non-trivial network load but realistic for field-level smart grid devices.

\subsubsection{Computational Capacity Model}
The compute capacity of each tier is represented by $\mu_n$, with values chosen to reflect the relative scarcity or abundance of resources: (a) \emph{Cloud tier}: $\mu_n = 60$ tasks/s, representing virtually unlimited compute resources of centralised data centres. (b) \emph{Fog tier}: $\mu_n = 15$ tasks/s, reflecting the moderate capacity of regional controllers (e.g., microgrid-level servers). c) \emph{Edge tier}: $\mu_n = 5$ tasks/s, highlighting the severe resource constraints of embedded IoT and DER-connected devices. To verify tractability, these values are scaled maintaining a proportional relationship of $\approx 12:3:1$ across tiers, consistent with reported resource hierarchies in hierarchical computing systems \cite{etsiMEC003}.

\subsubsection{Energy Consumption Model}
Energy consumption is modelled separately for processing and communication: (a) \emph{Processing energy per bit ($\eta_{\text{proc}}$)}: 2.0 J/bit (cloud), 0.5 J/bit (fog), 0.2 J/bit (edge). These values capture the inefficiency of large-scale cloud infrastructure compared to lightweight embedded devices optimised for specific tasks. (b) \emph{Communication energy per bit ($\eta_{\text{comm}}$)}: 1.2 J/bit (cloud), 0.3 J/bit (fog), 0.1 J/bit (edge). This reflects increasing transmission cost with distance: edge communication is local and cheap, fog is regional, and cloud requires long-haul wide-area data transport. The distinction between $\eta_{\text{proc}}$ and $\eta_{\text{comm}}$ enables clear attribution of energy costs to computation and data transport.

\subsubsection{Latency–Energy Tradeoff Weights}
The hybrid task assignment algorithm uses a composite cost function: $\Phi_{i,n} = \omega_1 L_{i,n} + \omega_2 E_{i,n},$ where $L_{i,n}$ is latency and $E_{i,n}$ is energy consumption. We select $\omega_1 = 0.6$ and $\omega_2 = 0.4$, prioritising latency-sensitive operation (critical for grid fault management and demand response) while still penalising energy-inefficient allocations. This reflects the real-world requirement that timely control actions are more critical than absolute energy minimisation.

\subsubsection{Simulation Scenarios}
Three scenarios are defined for comparative evaluation: \textit{Cloud-only baseline:} All tasks are processed at the cloud. This represents legacy centralised systems and highlights bottlenecks in latency and communication energy. \textit{Edge-first allocation:} Tasks are assigned preferentially to edge devices until the capacity is saturated, and then they overflow to fog and cloud. This represents a heuristic best-case for bandwidth reduction, but without optimisation. \textit{Greedy heuristic allocation:} Tasks are dynamically assigned to minimise the composite cost $\Phi_{i,n}$ using the proposed algorithm. This represents the optimised hybrid architecture.

\subsubsection{Simulation Methodology}
The simulation follows a repeatable pipeline: (1) \textbf{Task generation:} At each timestep, tasks are stochastically generated at edge nodes following the inter-arrival distribution defined by $\beta_i$. (2) \textbf{Feasibility filtering:} For each task, feasible computing nodes are identified based on latency constraints ($L_{i,n} \leq L_{\max}$) and available compute capacity ($\mu_n$). (3) \textbf{Cost computation:} For each feasible node, the composite cost $\Phi_{i,n}$ is calculated using latency and energy parameters. (4) \textbf{Task assignment:} This depends on the chosen scenario. (i) In the cloud-only case, all tasks are directed to the cloud. (ii) In the edge-first case, tasks are sequentially allocated to edge, fog, and then cloud until tier capacities are saturated. (iii) In the heuristic case, the tasks are assigned to the node with minimum $\Phi_{i,n}$. (5) \textbf{Resource update:} The capacity $\mu_n$ of each node is updated to reflect the assigned tasks. (6) \textbf{Metrics collection:} Energy consumption, average latency, bandwidth utilisation, and node utilisation are logged for analysis.

\subsubsection{Rationale for Parameter Choices}
The selected parameter values are designed to balance realism with tractability: a) The task rate (0.5–1.5 s) reflects variability in sensor-driven smart grid environments. b) The capacity scaling ratio (12:3:1) mirrors practical cloud–fog–edge hierarchies. c) Energy coefficients are consistent with reported differences in communication distances and processing overheads across tiers. d) Trade-off weights (0.6, 0.4) prioritise latency-critical performance while retaining energy awareness, consistent with smart grid operational priorities.

\subsection{Simulation Results}

Having established the simulation environment and parameter choices in the previous section, we now present the results obtained under different task allocation strategies and communication conditions. The objective of this analysis is to evaluate how the proposed greedy heuristic algorithm, when integrated into a hybrid edge–fog–cloud architecture, improves system performance relative to baseline approaches such as cloud-only, round-robin, or edge-first allocation. The results are organised thematically to reflect the different performance dimensions of the system. We begin by examining how tasks are distributed across computing tiers and how this affects the utilisation of edge, fog, and cloud resources. We then turn to the impact on bandwidth consumption, highlighting the effect of local preprocessing in reducing communication overheads. Next, we quantify energy efficiency gains achieved through intelligent offloading, followed by an evaluation of latency performance under both standard and URLLC-enabled conditions. The subsequent section investigates load balancing across computing tiers, demonstrating how dynamic allocation prevents bottlenecks and improves resilience. System-level stability and reliability are then analysed in terms of availability, curtailment, and packet loss under varying load and redundancy assumptions. Finally, we assess the role of distributed intelligence through AI-enabled fault detection and optimisation in community and microgrid use-cases. 

\subsubsection{Task Assignment and Resource Utilisation}

\begin{figure}[!t]
    \centering
    \includegraphics[width=0.50\textwidth]{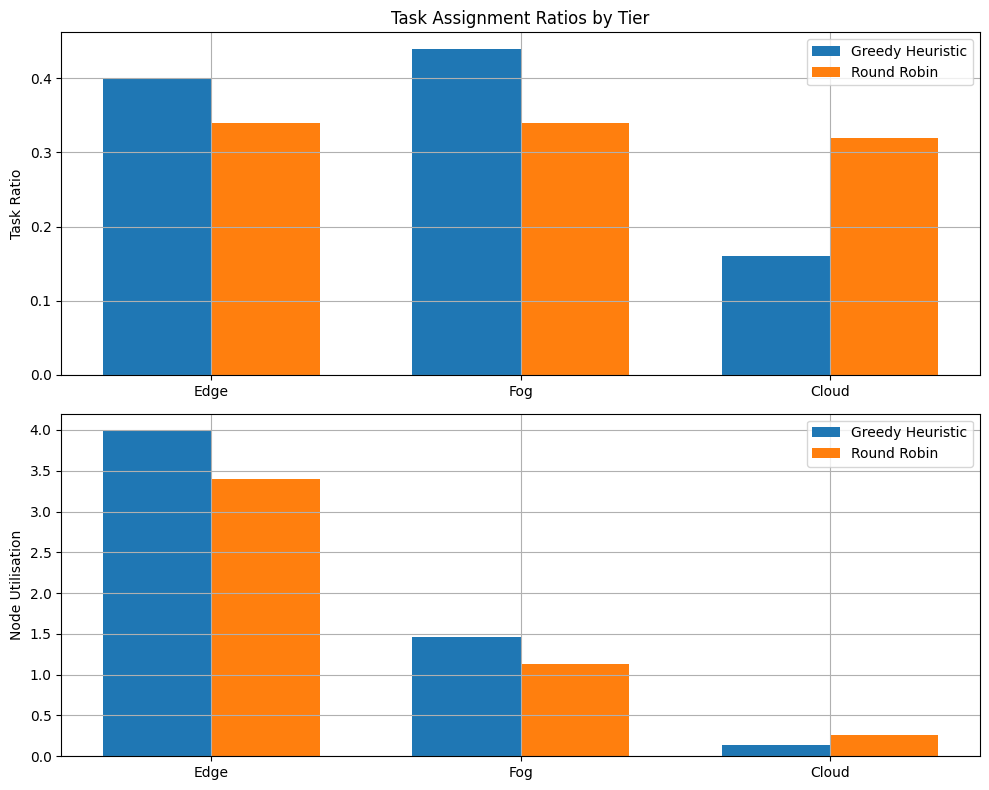}
    \caption{Task assignment ratios and node utilisation levels across edge, fog, and cloud tiers under greedy heuristic vs. round-robin allocation for $N=50$ tasks. Simulation setup: $\beta_i \in [0.67,2]$ tasks/s, $\mu_{\text{edge}}=5$, $\mu_{\text{fog}}=15$, $\mu_{\text{cloud}}=60$ tasks/s, $\eta_{\text{proc}}=\{0.2,0.5,2.0\}$ J/bit, $\eta_{\text{comm}}=\{0.1,0.3,1.2\}$ J/bit, $\omega_1=0.6$, $\omega_2=0.4$.}
    \label{fig:allocation}
    \vspace{-3mm}
\end{figure}

We first examine how tasks are distributed across computing tiers under different allocation strategies. When 50 tasks are generated and assigned using the greedy heuristic algorithm, the majority are offloaded to edge devices, with spillover directed to fog nodes, while only a minimal number reach the cloud. This contrasts with a round-robin allocation strategy, which distributes tasks evenly across tiers without regard for resource efficiency. As shown in Fig.~\ref{fig:allocation}, the greedy heuristic makes better use of resource-constrained but energy-efficient edge devices, while significantly reducing reliance on the cloud, which is both energy-intensive and distant. Physically, this behaviour emerges because the algorithm explicitly minimises a latency–energy cost, naturally steering tasks toward local, low-latency execution whenever capacity permits.

\subsubsection{Bandwidth Utilisation}

\begin{figure}[!t]
    \centering
    \includegraphics[width=0.50\textwidth]{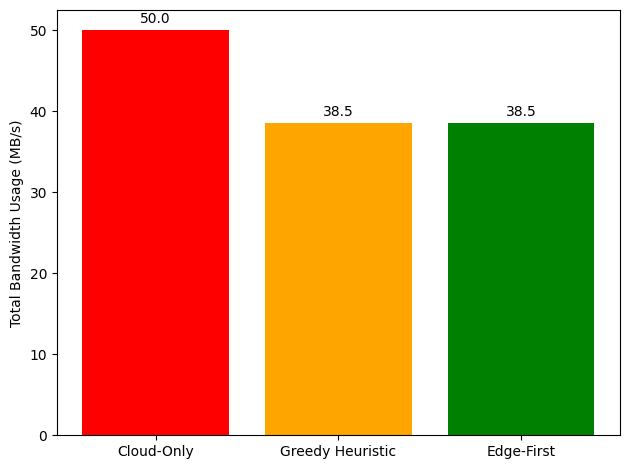}
    \caption{Bandwidth utilisation (Mbps) across cloud-only, edge-first, and greedy heuristic allocation models for $N=50$ tasks. Each task requires 1 MB/s bandwidth; preprocessing reduces transmission by $90\%$ at edge and $50\%$ at fog. Simulation setup: $\beta_i \in [0.67,2]$ tasks/s, $\mu_{\text{edge}}=5$, $\mu_{\text{fog}}=15$, $\mu_{\text{cloud}}=60$.}
    \label{fig:bandwidth}
\end{figure}

We next analyse the impact of task allocation on bandwidth usage. Figure~\ref{fig:bandwidth} compares three scenarios: cloud-only, edge-first, and greedy heuristic allocation. In the cloud-only scenario, all tasks are transmitted over the backbone network, resulting in the highest bandwidth demand. By contrast, edge-first allocation reduces bandwidth consumption by up to 90\% due to local preprocessing, while fog allocation offers a 50\% reduction. The greedy heuristic achieves nearly the same performance as edge-first, despite not hard-prioritising edge execution. This outcome indicates that the algorithm implicitly exploits bandwidth savings by avoiding unnecessary long-haul transmission. Physically, this arises because processing closer to the source filters and aggregates raw data, thereby suppressing redundant transmissions to higher tiers.

\subsubsection{Energy Efficiency}

\begin{figure}[!t]
    \centering
    \includegraphics[width=0.50\textwidth]{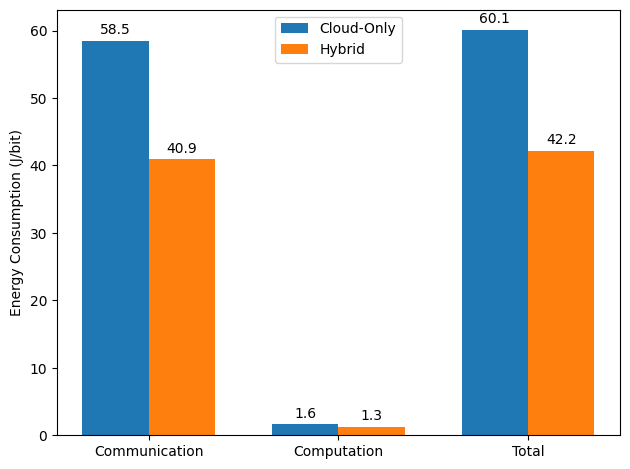}
    \caption{Comparison of total energy consumption (kWh) between cloud-only and hybrid edge–fog–cloud models under greedy heuristic allocation. Results averaged over 1000 trials with random seeds; maximum saving $69.65\%$, median saving $30.2\%$. Simulation setup: $N=50$ tasks, $\beta_i \in [0.67,2]$ tasks/s, $\eta_{\text{proc}}=\{0.2,0.5,2.0\}$ J/bit, $\eta_{\text{comm}}=\{0.1,0.3,1.2\}$ J/bit, $\omega_1=0.6$, $\omega_2=0.4$.}
    \label{fig:energy}
\end{figure}

Energy consumption is then evaluated across centralised and hybrid architectures. Figure~\ref{fig:energy} shows that offloading tasks using the greedy heuristic substantially reduces total energy use compared to a cloud-only approach. The reduction stems from both shorter transmission paths, which cut communication energy, and more efficient edge processors, which lower computational energy. Running the trial 1000 times reveals a maximum saving of 69.65\%, with a 30.2\% median. Occasional zero savings occur when all tasks are directed to the cloud due to capacity or latency constraints. This behaviour reflects the fact that cloud processing incurs high communication and processing costs, while edge and fog tiers deliver cheaper and closer computation when capacity allows.

\subsubsection{Latency and URLLC Support}

\begin{figure}[!t]
    \centering
    \includegraphics[width=0.9\textwidth]{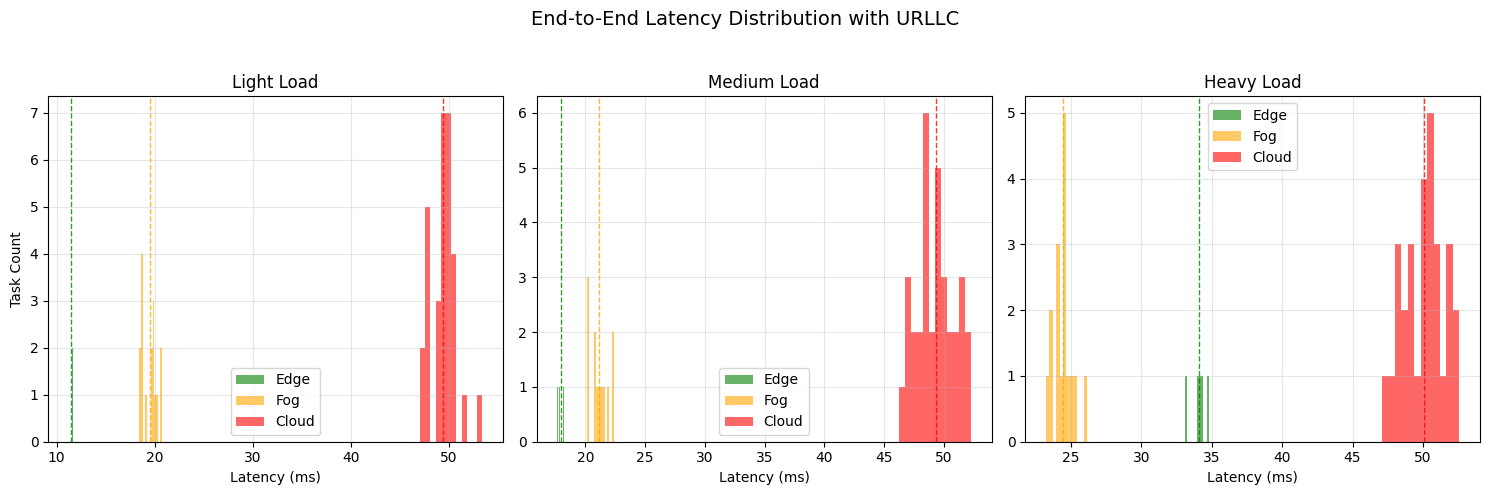}
    \caption{End-to-end latency distribution under greedy heuristic allocation with URLLC enabled. Transmission delay reduced by $15\%$ and jitter by $25\%$ relative to baseline. Simulation setup: varying workloads up to $N=100$ tasks, $\beta_i \in [0.67,2]$ tasks/s, $\mu_{\text{edge}}=5$, $\mu_{\text{fog}}=15$, $\mu_{\text{cloud}}=60$, $\omega_1=0.6$, $\omega_2=0.4$.}
    \label{fig:latency_with}
\end{figure}

\begin{figure}[!t]
    \centering
    \includegraphics[width=0.9\textwidth]{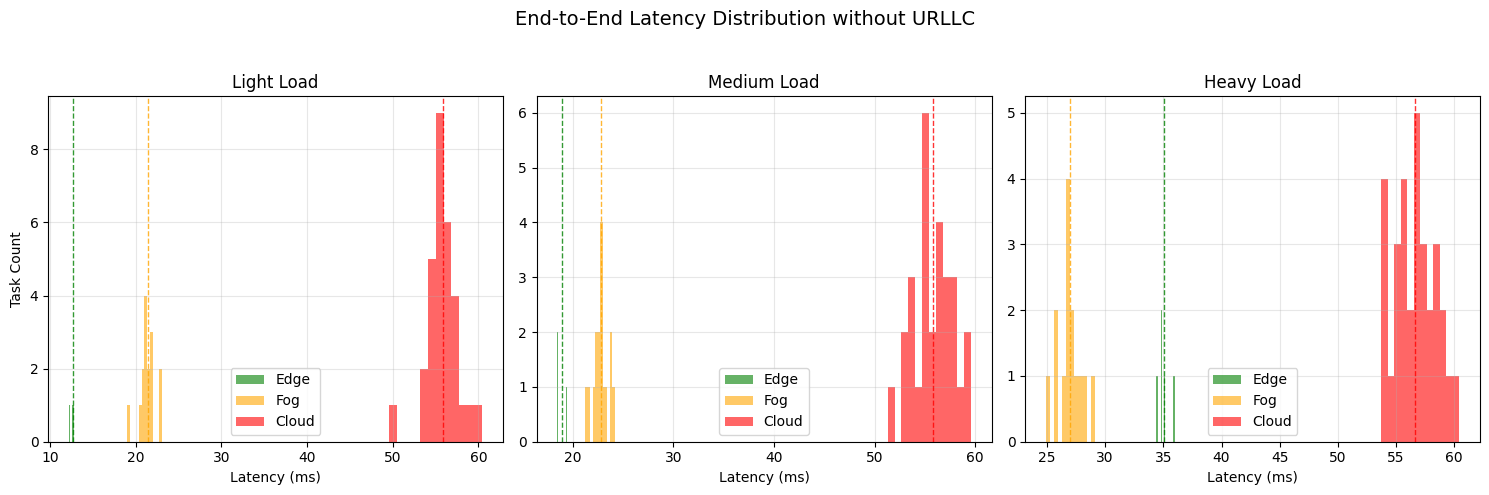}
   \caption{End-to-end latency distribution under greedy heuristic allocation without URLLC. Simulation setup identical to Fig.~\ref{fig:latency_with}, but without delay/jitter reduction.}
    \label{fig:latency_without}
\end{figure}

Latency performance was further studied under varying workloads, with and without URLLC enabled. Figures~\ref{fig:latency_with} and \ref{fig:latency_without} show that as load increases, more tasks are pushed from edge to fog nodes, introducing additional delay. The greedy heuristic ensures that latency remains bounded, but URLLC introduces further improvements: transmission delay is reduced by 15\% and jitter by 25\%. Physically, this occurs because URLLC ensures prioritised, reliable transmission with duplication and slicing mechanisms, allowing time-sensitive tasks to be delivered more consistently. The combination of heuristic allocation and URLLC thus enhances the system’s ability to support real-time grid operations.

\subsubsection{Load Balancing and Resource Utilisation}

\begin{figure}[!t]
    \centering
    \includegraphics[width=0.7\textwidth]{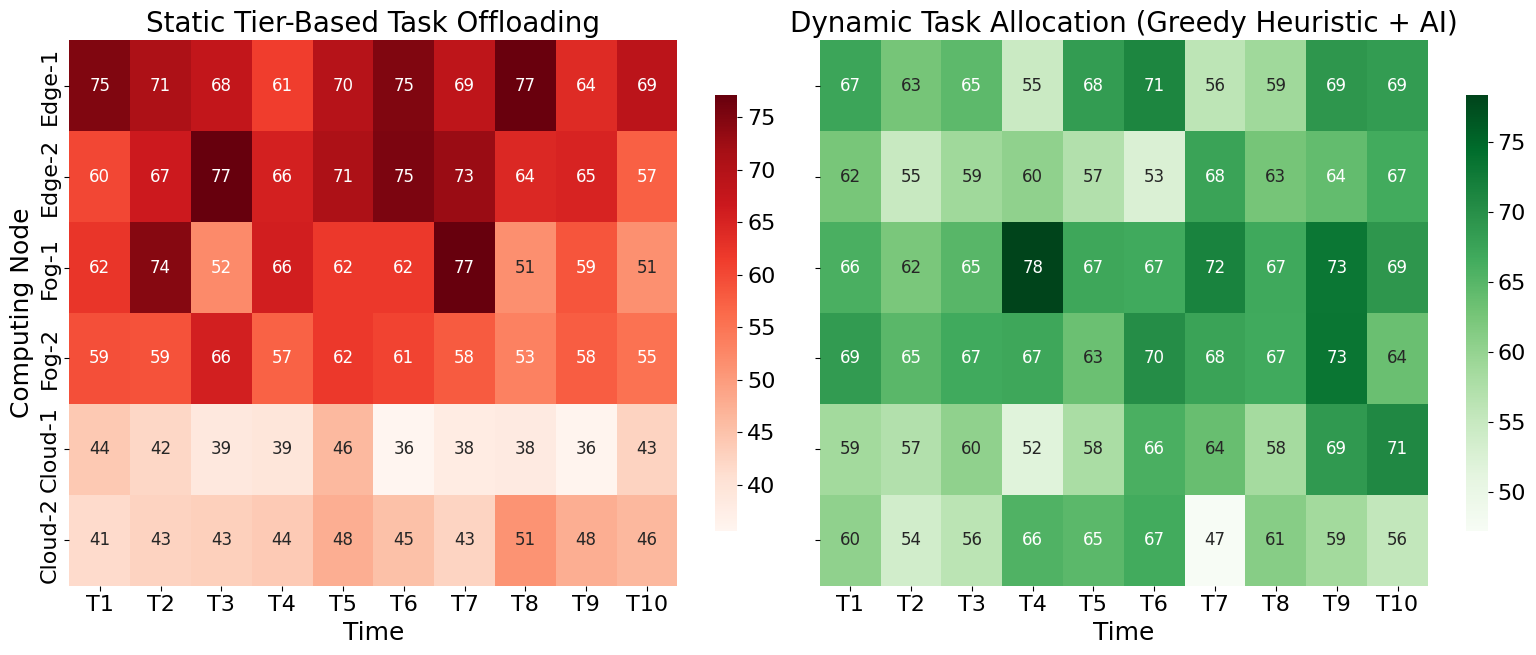}
    \caption{CPU utilisation heatmap of computing nodes under static round-robin allocation vs. dynamic greedy heuristic allocation with AI predictions for $N=50$ tasks. Simulation setup: $\beta_i \in [0.67,2]$ tasks/s, $\mu_{\text{edge}}=5$, $\mu_{\text{fog}}=15$, $\mu_{\text{cloud}}=60$, $\eta_{\text{proc}}=\{0.2,0.5,2.0\}$ J/bit, $\eta_{\text{comm}}=\{0.1,0.3,1.2\}$ J/bit.}
    \label{fig:heatmap}
\end{figure}

Figure~\ref{fig:heatmap} compares CPU utilisation across computing nodes under static round-robin allocation and dynamic heuristic allocation. In the static case, edge devices are consistently overloaded while cloud resources remain underused. This imbalance results because tasks are naively distributed without regard for complexity, capacity, or energy–latency trade-offs. In the dynamic case, tasks are more evenly distributed, preventing bottlenecks at the edge and leveraging fog and cloud capacity more effectively. The observed behaviour highlights the importance of intelligent allocation in smoothing utilisation across tiers, thereby improving resilience and avoiding single-tier overload.

\subsubsection{System Stability and Reliability}

\begin{figure}[!t]
    \centering
    \includegraphics[width=0.60\textwidth]{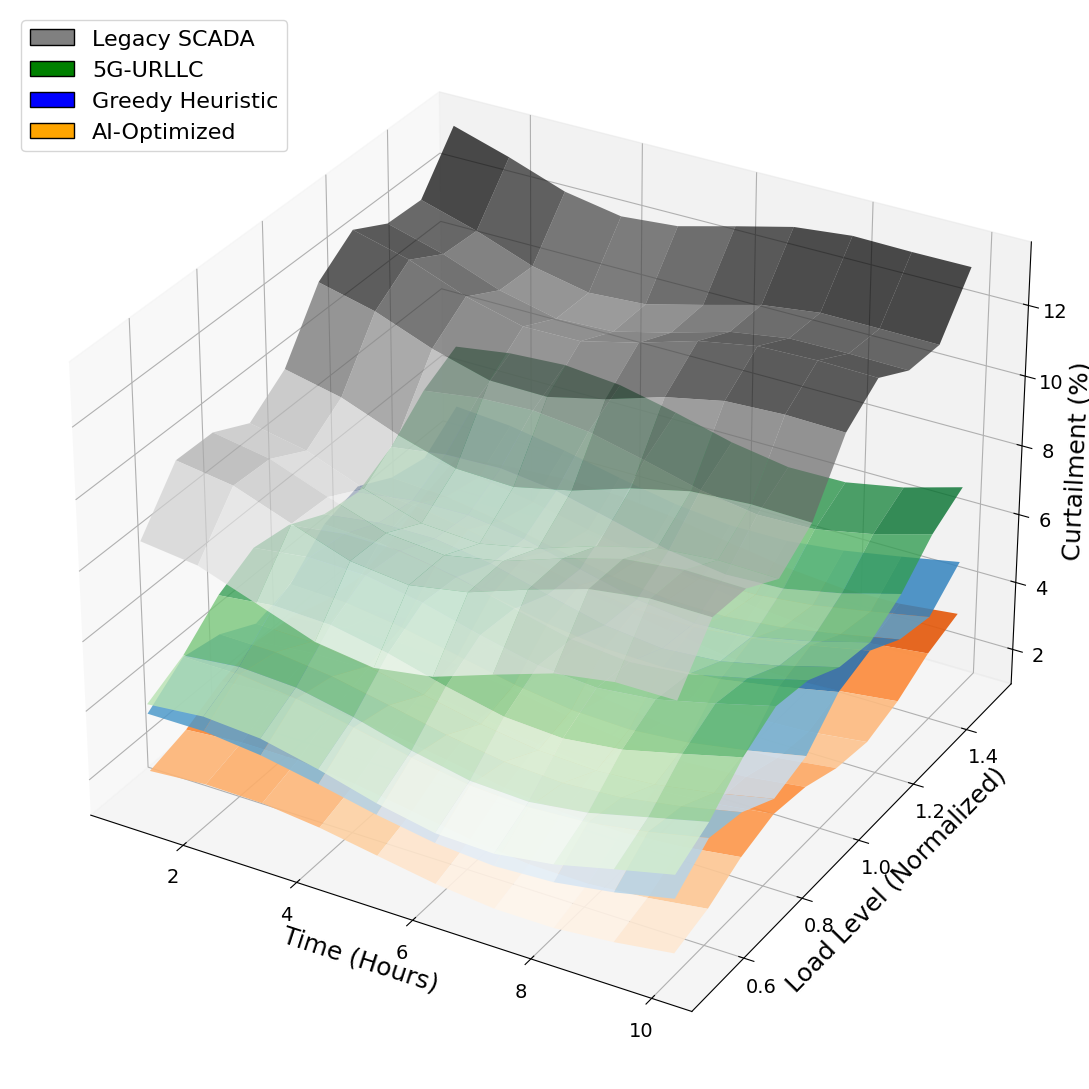}
    \caption{Percentage of generation curtailment under centralised SCADA, URLLC-enabled, heuristic allocation, and AI-optimised coordination. Simulation setup: $N=100$ tasks, $\beta_i \in [0.67,2]$ tasks/s with stochastic DER output variability.}
    \label{fig:curtailment}
\end{figure}

\begin{figure}[!t]
    \centering
    \includegraphics[width=0.60\textwidth]{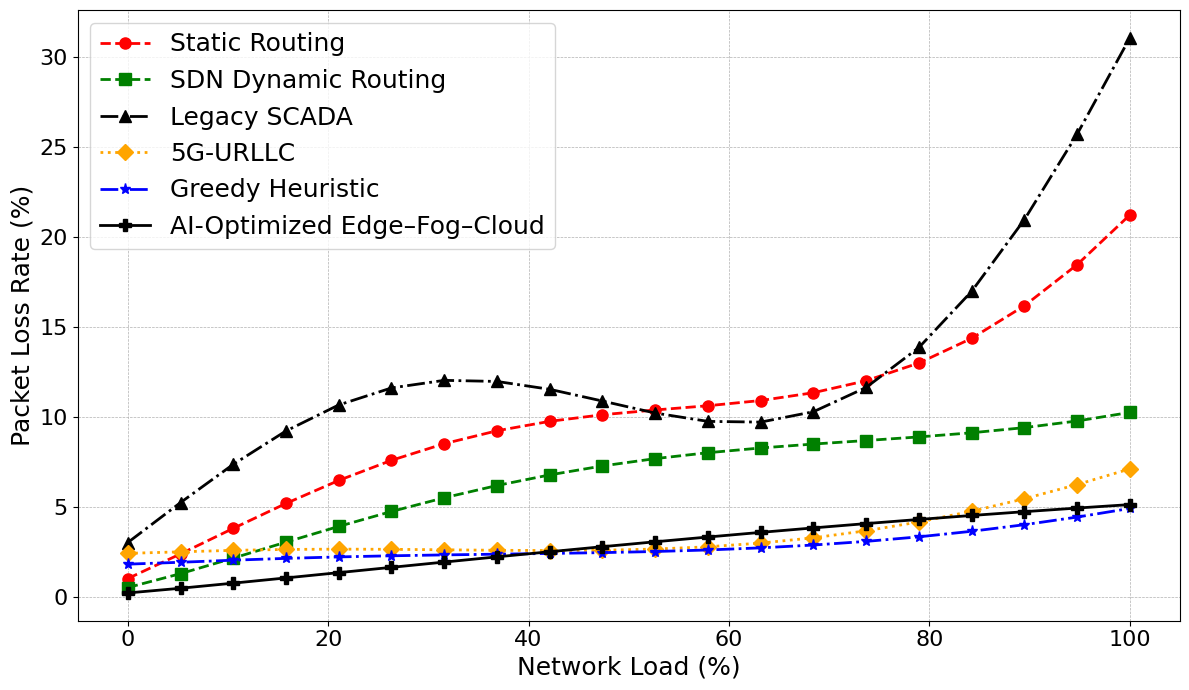}
    \caption{Packet loss rate (\%) versus network load under static routing, SDN, URLLC, heuristic allocation, and AI optimisation. Simulation setup: network load varied up to 95\% channel capacity, $N=100$ tasks, $\beta_i \in [0.67,2]$ tasks/s.}
    \label{fig:packetloss}
\end{figure}

At the system level, stability and availability were evaluated. Using the availability model in (1), a centralised system with one-hour repair time and one-year mean-time-to-failure achieves 99.989\% availability. By contrast, the hybrid system, which introduces redundancy through SDN rerouting and multi-tier allocation, raises availability to 99.9999999998466\%, equivalent to only 48~$\mu$s downtime per year (Table~\ref{tab:availability}). This dramatic improvement arises physically because failures are no longer single points of failure: tasks can be reassigned to alternative routes or tiers.

\begin{table}[!t]
\footnotesize
\renewcommand{\arraystretch}{1.2}
\caption{Service availability comparison.}
\label{tab:availability}
\centering
\begin{tabular}{|l|c|c|}
\hline
\textbf{Metric} & \textbf{Centralised (single point)} & \textbf{Hybrid ($>3$ layers)} \\
\hline
Availability & 99.9884673\% & 99.9999999998466\% \\
MTTF & 8670h & $6.57 \times 10^{11}$h \\
Downtime/year & 1h & 48~$\mu$s \\
\hline
\end{tabular}
\end{table}

We also examine curtailment and packet loss. Figure~\ref{fig:curtailment} shows that centralised SCADA-based systems suffer high curtailment under load, as delayed responses prevent optimal use of DER output. URLLC-based communications reduce curtailment, and integrating heuristic allocation further lowers it by enabling local decision-making. AI optimisation delivers the lowest curtailment, owing to predictive scheduling. Similarly, packet loss (Fig.~\ref{fig:packetloss}) is much higher in static routing, particularly at saturation. SDN improves packet delivery through adaptive routing, URLLC adds reliability through duplication and slicing, and heuristic plus AI approaches further reduce losses by anticipating congestion. These results highlight that hybrid architectures improve both energy yield and communication quality by reducing sensitivity to load and failure.

\subsubsection{Intelligence-Enabled Performance}

\begin{figure}[!t]
    \centering
    \includegraphics[width=0.6\columnwidth]{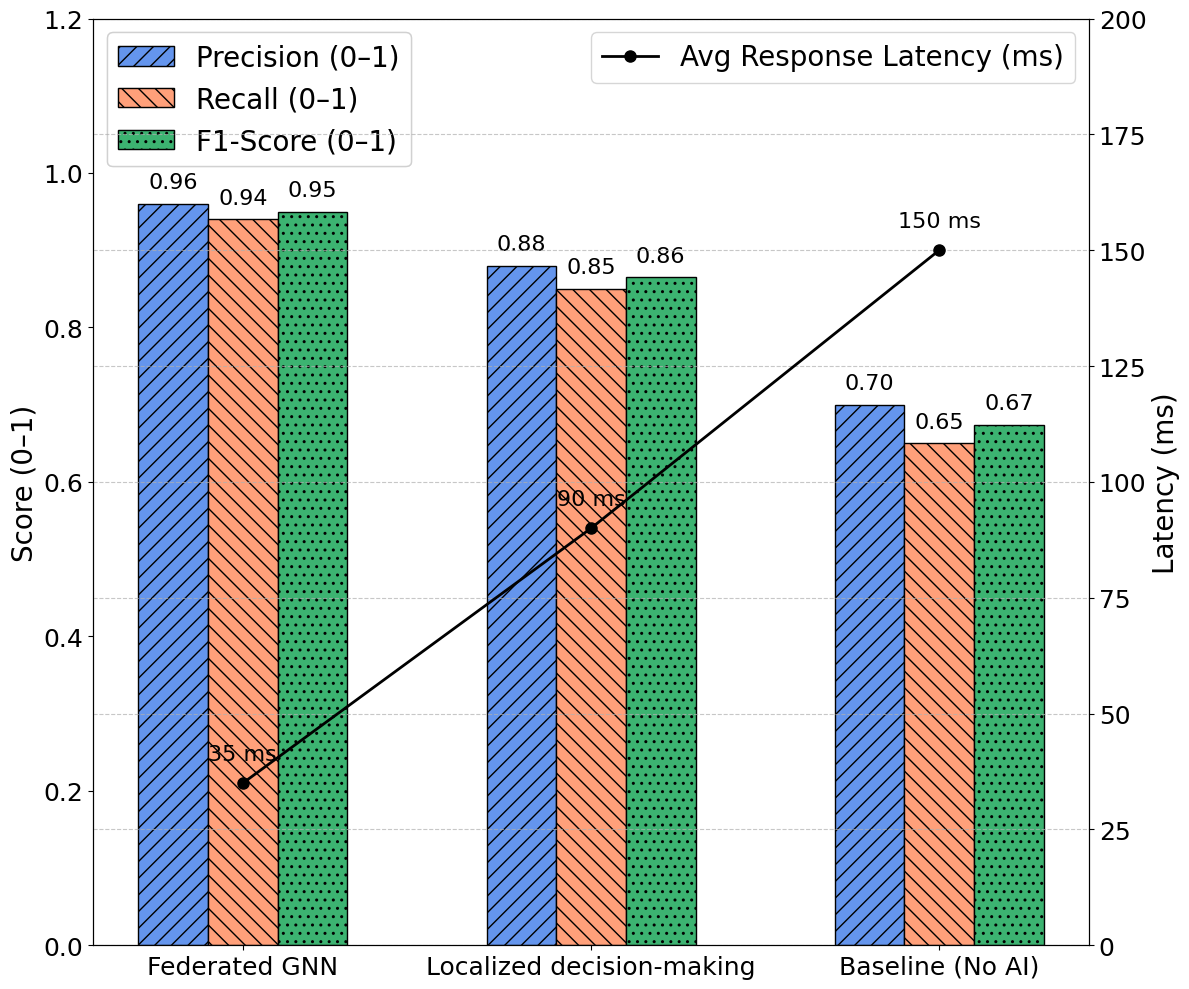}
    \caption{Fault detection performance of baseline thresholding, local edge–fog decision-making, and federated GNN. Metrics: precision, recall, F1-score, and average response latency. Simulation setup: $N=50$ tasks, $\beta_i \in [0.67,2]$ tasks/s, federated GNN trained on decentralised graph data across tiers.}
    \label{fig:fault}
\end{figure}

Finally, we evaluate the role of distributed intelligence. Figure~\ref{fig:fault} compares three approaches to fault detection: a baseline threshold-based system, a localised edge–fog model, and a federated GNN implementation. The baseline system achieves low accuracy (F1 = 0.67) and high latency (150 ms) due to reliance on delayed centralised control. Localised edge–fog intelligence improves both accuracy (F1 = 0.86) and latency (90 ms), but the federated GNN achieves the best results, with precision of 0.96, recall of 0.94, F1 = 0.95, and latency of 35 ms. Physically, this is explained by the fact that federated GNNs exploit global patterns without transmitting raw data, thereby preserving both responsiveness and scalability.

\begin{figure}[!t]
    \centering
    \includegraphics[width=0.6\columnwidth]{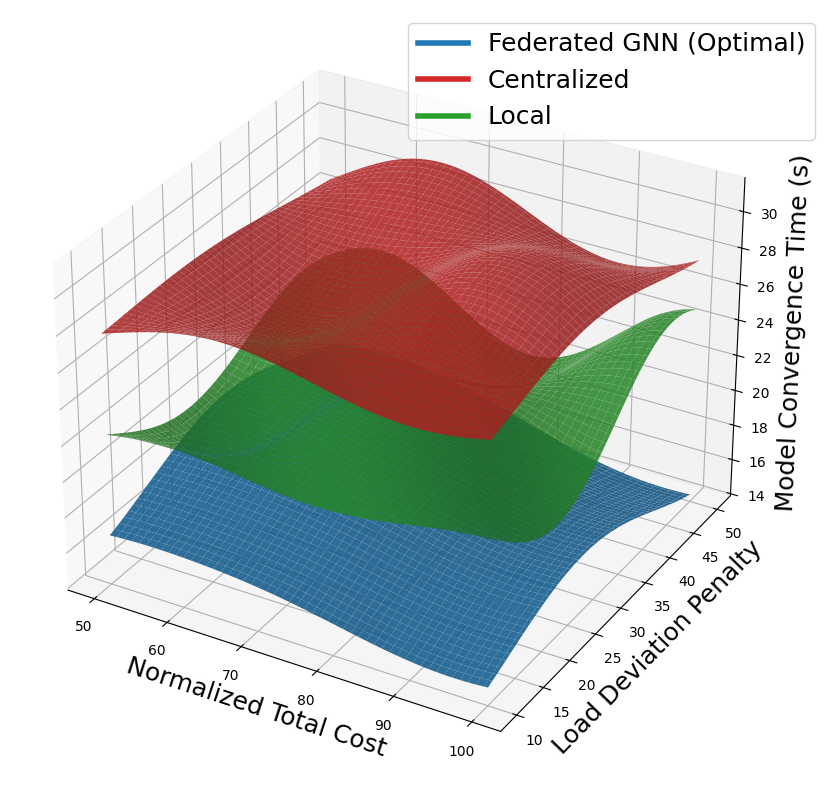}
    \caption{Performance of centralised optimisation, local-only learning, and federated GNN in community energy coordination. Metrics: convergence time, cost, and load deviation penalties. Simulation setup: $N=10$ microgrids, $K=100$ IoT nodes, $\beta_i \in [0.67,2]$ tasks/s, $\mu_{\text{edge}}=5$, $\mu_{\text{fog}}=15$, $\mu_{\text{cloud}}=60$.}
    \label{fig:community}
\end{figure}

Distributed AI is also tested for energy community coordination (Fig.~\ref{fig:community}). Centralised optimisation suffers from scalability limits, while local-only learning is faster but inconsistent under complexity. The federated GNN consistently converges below 15 seconds, even under increasing load deviation penalties, while maintaining high accuracy. This reflects the ability of federated learning to combine global coordination with local responsiveness, while preserving privacy by avoiding raw data sharing. Physically, this enables distributed communities and microgrids to operate autonomously, yet remain coordinated for system-wide efficiency.

\section{Conclusion} \label{sec:conclusions}

In this paper, we presented the first comprehensive exploration of a SDEN that integrates edge, fog, and cloud resources through an energy–latency–aware task offloading heuristic for smart grid applications. Unlike prior work that has typically focused on isolated aspects such as cloud-only optimisation or fog-assisted scheduling, we demonstrated for the first time how a fully hybrid architecture, jointly optimised with modern 5G features (URLLC, SDN, NFV) and distributed AI, can deliver simultaneous gains in latency, energy efficiency, bandwidth reduction, reliability, and fault detection.

These results collectively mark the first demonstration that a hybrid SDEN architecture, paired with lightweight optimisation and distributed intelligence, can address multiple pain points in smart grid operation—bandwidth overheads, energy inefficiency, latency sensitivity, and reliability constraints—within a single integrated framework. As future work, we plan to extend this novel framework into hardware-in-the-loop and field trials, validate interoperability with DERMS and legacy SCADA systems, harden security and resilience against adversarial conditions, and explore carbon- and price-aware scheduling that co-optimises ICT and grid objectives. By advancing both the architecture and the intelligence that drive it, we aim to translate these first-of-their-kind simulation results into practical, scalable deployments for next-generation smart grids.

In the future, we will investigate cases where the nodes within a tier are heterogenous, and we will try to address uncertainties and dynamic load arrivals.



\bibliographystyle{IEEEtran}
\bibliography{ref}

\end{document}